\documentclass[aps,pra,twocolumn,groupedaddress]{revtex4-1}

\usepackage{amsmath}
\usepackage{amsthm}
\usepackage{amssymb}
\usepackage{graphicx,hyperref,bbm,enumitem,bm}  
\usepackage[pdftex]{color}

\usepackage[toc,page]{appendix}

\def\ave#1{\langle #1 \rangle}
\def\ii{{\rm i}}
\def\sx{\sigma^{\rm x}}
\def\sy{\sigma^{\rm y}}
\def\sz{\sigma^{\rm z}}
\def\tr#1{{\rm tr}{#1}}
\def\1{\mathbbm{1}}
\def\ket#1{{| #1 \rangle}}
\def\bra#1{{\langle #1 |}}

\def\ax{a_{\rm x}}
\def\ay{a_{\rm y}}
\def\az{a_{\rm z}}
\def\tc{t_{\rm c}}
\def\rA{\rho_{\rm A}}

\def\tit#1{{\em #1},}

\newcommand{\new}[1]{{#1}}

\begin{document}
	
\title{Two-step relaxation in local many-body Floquet systems}
	
	\author{Marko \v Znidari\v c}
	\affiliation{Department of Physics, Faculty of Mathematics and Physics, University of Ljubljana, 1000 Ljubljana, Slovenia}
	
	\date{\today}
	
	\begin{abstract}
         We want to understand how relaxation process from an initial non-generic state proceeds towards a long-time typical state reached under unitary quantum evolution. One would expect that after some initial correlation time relaxation will be a simple exponential decay with constant decay rate. We show that this is not necessarily the case. Studying various Floquet systems with fixed two-qubit gates, and focusing on purity and out-of-time-ordered correlation functions, we find that in many situations relaxation proceeds in two phases of exponential decay having different relaxation rates. Namely, in the thermodynamic limit the relaxation rate exhibits a change at a critical time proportional to system's size. \new{The initial thermodynamically relevant rate can be slower or faster than the asymptotic one, demonstrating that the recently discovered phantom relaxation, in which the decay is slower than predicted by a nonzero transfer matrix gap, is not limited to only random circuits.}
	\end{abstract}
	
	\maketitle

\section{Introduction}
        
        Complexity of dynamics in general increases with increasing number of particles. However, when the number of particles gets very large ``self-averaging'' can come to our rescue and certain properties can again become simpler. In the thermodynamic limit almost any state at given fixed values of conserved quantities from a large Hilbert space will have the same expectation values of sufficiently well behaved observables. While this makes such typical states rather dull and featureless they are crucial for understanding why statistical physics works so well and is so general~\cite{landau}. Properties of typical states are rather well studied and understood. To get their properties one can employ different techniques, either from quantum chaos and random matrix theory~\cite{Haake,felix}, simply use an appropriate equilibrium statistical ensemble~\cite{landau}, use measure concentration techniques~\cite{neumann,Hayden,joel}, or, being careful about the order of time $t \to \infty$ and system size $n \to \infty$ limits, use expansion into eigenstates~\cite{eth}. Often though we are interested in non-typical states, for instance, the initial state undergoing quantum evolution might be special. While we can immediately say that for generic quantum evolution such non-typical initial state will eventually evolve into a typical one, much less is known about the process of relaxation from the initial non-typical state to the long-time stationary behavior of typical states~\cite{anatoli,gogolin}. In the present paper we are going to reveal new surprising property of such relaxation in local Floquet systems composed of non-commuting nearest-neighbor gates. 

\begin{figure}[t!]
\centerline{\hskip1cm  \includegraphics[height=1.02in]{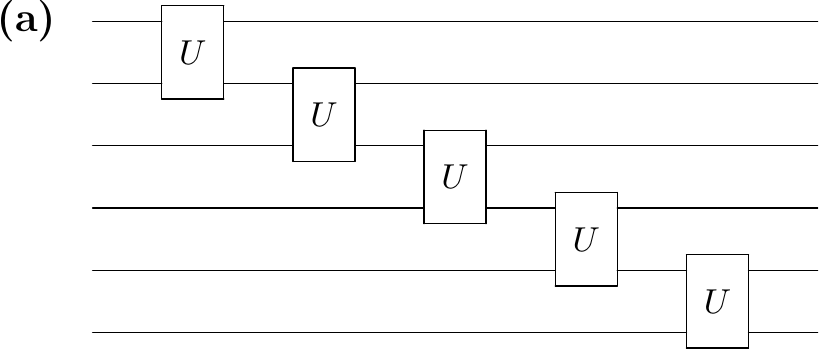}\hfill}
  \vskip5mm
\centerline{\hskip1cm   \includegraphics[height=1.36in]{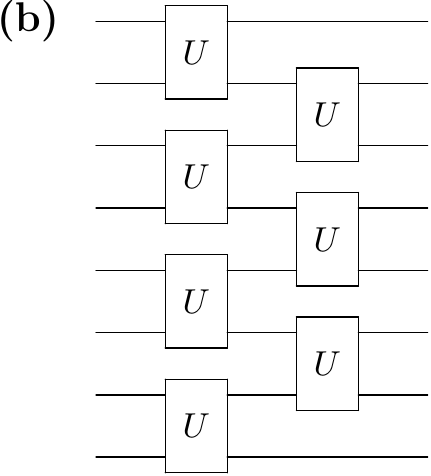}\hfill}
\caption{Circuit configurations with open boundaries studied (time increases from left to right). Frame (a) shows the staircase (S), while (b) is the brick-wall (BW) configuration showing a one-step Floquet propagator $U(1)$ that is made of the same two-qubit gate $U$.}
\label{fig:conf}
\end{figure}

Time-periodic or so-called Floquet systems have a long history. In single-particle quantum chaos they are widely used as one of the simplest settings that can display chaos, a famous example being the kicked rotator (quantized standard map)~\cite{casati}. More recently motivation comes from trying to understand many-body physics as well as from quantum information. There are multiple reasons to consider Floquet systems. Sometimes they are easier to understand; writing a single-step propagator $U(t=1)=V_1 V_2$ as a product of two propagators $V_{1,2}$, like in the kicked rotator where $V_1$ is the kinetic term and $V_2$ the potential, can result in rich physics even when each constituent $V_{1,2}$ alone have simple non-chaotic dynamics. Continuing this splitting idea one can write $U(1)$ in terms of individual two-body gates as done in quantum computation, resulting in a quantum circuit (Fig.~\ref{fig:conf}). Such formulation is also extremely handy for modern numerical simulation methods based on the matrix product ansatz~\cite{ulrich}. In the many-body context a number of such circuits has been recently found to be solvable. Some examples are solvable random circuits~\cite{chalker18,adam18,vedika,tianci,markov,skin22}, dual unitary circuits~\cite{brunoprl19,bruno19,bruno23}, and other~\cite{sarang,katja,lamacraft21}. Not least, Floquet formulation in terms of individual quantum gates is also natural in certain experimental settings like ultracold gases~\cite{bloch} or quantum computers~\cite{google}.

Motivation for our study comes from recent discovery that in certain random circuits relaxation as measured by entanglement or out-of-time-ordered correlation (OTOC) functions does not proceed in a simple exponential manner~\cite{prx21,otoc22,bipart}. \new{Rather, relaxation proceeds in two phases, each displaying exponential decay, but with different relaxation rates that changes in the thermodynamic limit at an extensive time.} While physical mechanism being at work is not yet fully understood, more clarity was provided by a solvable case of the staircase configuration of random two-qubit gates (each gate is an independent random unitary)~\cite{skin22}. Namely, the effect can be traced back~\cite{skin22,skin23} to non-Hermiticity of the underlying Markovian average dynamics: left and right eigenvectors localize at the edge (a.k.a. the non-Hermitian skin effect~\cite{skin1,skin2}), causing exponentially growing spectral expansion coefficients. On the level of e.g. purity evolution this is then reflected in the fact that the initial thermodynamically relevant relaxation rate is not given by the 2nd largest eigenvalue, as one would expect, but rather by the pseudospectrum~\cite{trefethen}. In a nutshell, having Markovian description with a finite spectral gap the 2nd largest eigenvalue does not necessarily determine the correct relaxation rate. Somewhat similar observations of discrepancy between the spectrum and the relaxation rate have been recently observed in other situations that involve non-Hermitian matrices, see Refs.~\cite{song,takashi20,takashi21,ueda21,lorenza21,clerk}.

Common to all of the above situations is non-Hermiticity, and in random circuits also the explicit randomness -- all gates are different and statistically independent. An obvious question is whether the two-step relaxation is specific to random circuits, or, can it occur more generally? Also, is an exact non-Hermitian description, like in the mentioned cases, necessary for the phenomenon? By studying Floquet circuits we show that the two-step relaxation is more widespread: (i) no randomness is required -- evolution can be fully deterministic and homogeneous, and (ii) no explicit non-Hermitian matrices need to be involved -- an effective non-Hermiticity will come in automatically as soon as one considers e.g. local observables, or performs a partial trace, so that the coherence of full unitary evolution is ``lost''.

\new{Note that the main point is not that the relaxation is not a single exponential -- there are plenty of well known situations in physics with non-exponential relaxation (e.g., inhomogeneous systems where one has a distribution of relaxation rates, systems where the gap closes in the thermodynamic limit, etc.) -- but rather that in such two-step phantom relaxation the rate is not equal to what one expects it to be (the gap). In short, relaxation is still exponential but with seemingly ``wrong'' rate. The phenomenon is also not related to the so-called prethermalization in which relaxation is delayed due to high-frequency or high-strength driving, see Ref.~\cite{Ho23} for a review.}

\section{Floquet evolution}

We are going to study relaxation of an $n$ qubit system under unitary Floquet evolution. The main quantity that we shall look at will be purity $I(t)$,
\begin{eqnarray}
      \label{eq:I}
  I(t)&=&\tr{\rA^2(t)}, \qquad \rA(t)={\rm tr}_{\rm B} \ket{\psi(t)}\bra{\psi(t)},\\
  \ket{\psi(t)}&=&U(t)\ket{\psi(0)}, \qquad U(t)=U(1)^t,\nonumber
    \end{eqnarray}
where the unitary propagator $U(t)$ for an integer $t$ is a simple power of a single-step Floquet propagator $U(1)$. The Floquet propagator for one unit of time $U(1)$ is made out of same two-qubit gates $U_{k,k+1}$ applied to different neighboring qubits in either a staircase configuration (abbreviated by S), see Fig.~\ref{fig:conf}(a), or in a brick-wall configuration, see Fig.~\ref{fig:conf}(b). The two-qubit gate $U$ will have an XXZ-like form that will be specified latter. We shall vary two types of boundary conditions, either open boundary conditions (OBC) shown in Fig.~\ref{fig:conf}, or periodic boundary conditions (PBC) where the final $n$th gate in $U(1)$ is applied between the last and the first qubit, i.e., $U(1)_{\rm PBC}=U_{n,1} U(1)_{\rm OBC}$. The total number of two-qubit gates in $U(t)$ is therefore $(n-1)t$ for OBC and $tn$ for PBC. The initial state $\ket{\psi(0)}$ will always be a random product state, $\psi(0)=\prod_{k=1}^n \otimes \chi_k$, where $\chi_k$ are independent random single-qubit states. Purity measures bipartite entanglement and is for our separable initial states $I(0)=1$, after which it starts to decay. The smaller the purity the larger bipartite entanglement there is in $\psi(t)$. We will always use a half-half bipartition where the subsystem A consists of the 1st $n/2$ qubits, and B the rest. We expect behavior to be similar also for other more complicated bipartitions, similar to the situation in random circuits~\cite{bipart}.

Because our Floquet propagator $U(1)$ will be generic (i.e., quantum chaotic according to level spacing statistics), and initial states are in no way special (e.g., eigenstates of $U(1)$), the state reached after long time will have the same properties as a typical random state. In particular, purity will asymptotically converge towards that of random states~\cite{karol01} $I_\infty=2 N_{\rm A}/(1+N_{\rm A}^2)$, where the subsystem size is $N_{\rm A}=2^{n/2}$. Because we want to study relaxation towards long-time stationary state we will always look at the decay of $I(t)-I_\infty$. For generic (chaotic) evolution one would expect that after some initial transient time, whose length does not scale with $n$, the decay will be exponential,
\begin{equation}
  I(t)-I_\infty \sim {\rm e}^{-r t},
  \label{eq:exp}
\end{equation}
with a rate $r$ that will depend in the chosen $U$ and configuration, but will be constant in time and will not depend on $n$ in the thermodynamic limit (TDL).

Our main result will be that $r$ in fact does change with time. As we will see, $r$ will change at a critical $\tc \sim n$ so that the rate will be equal to $r_{\rm I.}$ in the first relaxation phase, and to different $r_{\rm II.}$ in the second phase. Grossly speaking the phase II. describes relaxation when $\psi(t)$ is already close to being a random state, while the phase I. describes relaxation when $\psi(t)$ is still far from being random. If one takes the TDL (first taking the infinite system size limit and only then $t \to \infty$) the relevant relaxation phase is I.. In such TDL one will therefore never observe $r_{\rm II.}$. However, simply labeling the 2nd phase as being irrelevant would miss an important and interesting physics. Often one is interested in dynamics (relaxation) on time scales that are larger than linear in the system size, i.e., beyond $\tc$ upto which only $\sim n^2$ gates are applied. This might in particular be the case for quantum algorithms that require polynomial number of gates~\cite{nielsen}, e.g., Shor's factorization runs in $\sim n^3$ gates (i.e., in our units $t \sim n^2$).

It is instructive to first find the maximal possible rate at which entanglement can be produced for a given circuit configuration. From Fig.~\ref{fig:conf} we see that regardless of configuration being S or BW there is only one gate $U$ per unit of time that connects subsystems A and B (symmetric half-half bipartition) for OBC, while there are 2 for PBC. One two-qubit gate can increase entanglement by at most $2$ ebits~\cite{max}, i.e., increase the rank of the reduced density operator $\rA$ by at most $4=2^2$. For this increase to be really $2$ ebits (and not $1$) both qubits have to be entangled with other qubits in subsystems A and B, respectively.  If this is not the case the two-qubit $U$ can increase entanglement by at most 1 ebit, i.e., increase rank of $\rA$ by a factor of $2=2^1$. As an example, a SWAP gate acting on qubits 2 and 3 in a separable state $\ket{00+11}_{\rm A} \ket{00+11}_{\rm B}$ will increase the rank of $\rA$ from 1 to 4. On the other hand, if either the 2nd qubit is factorized from the 1st, or the 3rd from the 4th, e.g. $U$ acting on $\ket{00+11}_{\rm A} \ket{00}_{\rm B}$, the rank of $\rA$ can increase from 1 to at most 2. This is important for the distinction between S and BW configurations. In the BW circuit (see Fig.~\ref{fig:conf}(b)) the two qubits on which $U_{n/2,n/2+1}$ acts have already been acted on with previous gates $U$; therefore the rank of $\rA$ will increase by a factor of 4. In the S circuits, on the other hand, qubits in subsystem B (sites $n/2+1,\ldots n$) have not yet been touched since the last step the connecting gate $U_{n/2,n/2+1}$ has been applied. Therefore, the rank of $\rA$ will increase only by a factor of 2. One can use this to get the maximal rank after time $t$ for all circuits studied. For instance, for the S with OBC the rank is $2^t$, whereas for PBC where there are two gates connecting A and B per time step the maximal rank is $2^{2t+1}$ (an extra $1$ in the exponent comes due to first $U_{n,1}$ that acts on already entangled qubits). The maximal possible rank for all circuits studied is summarized in Table~\ref{tab:rank}. Such rank of course holds only until it reaches its maximal possible value, which is $2^{n/2}$ for our half-half bipartition. Except for a measure zero of initial states (e.g., eigenstates of $U(1)$) this maximal rank is also realized in actual evolutions we study. Entanglement would be maximal if all eigenvalues $\lambda_j$ of $\rA$ would be equal, in which case one would have a minimal $I(t)=\sum_j^{\rm rank} \lambda_j^2=1/({\rm rank})$. Such fastest possible purity decay would in turn give the largest possible relaxation rate $r_{\rm I.}$. For instance, for the BW-OBC one has the maximal rank $4^t$ and therefore the fastest possible purity decay has ${\rm e}^{-r_{\rm I.}t}=(1/4)^t$. Such maximal decay rates are actually saturated in random circuits with XXZ-type two-qubit gates~\cite{prx21}.

\begingroup
\squeezetable
\begin{table}[t!]
  \begin{ruledtabular}
  \begin{tabular}{lcc}
Configuration & rank of $\rA(t)$ & fastest possible decay of $I(t)$\\
\hline
BW OBC  & $2^{2t}$ ($2^{2t-1}$ for $n=4k+2$) & $(\frac{1}{4})^t$  \\
BW PBC  & $2^{4t}$ ($2^{4t-1}$ for $n=4k+2$) & $(\frac{1}{16})^t$ \\
S OBC  & $2^t$ & $(\frac{1}{2})^t$ \\
S PBC  & $2^{2t+1}$ & $(\frac{1}{4})^t$
\end{tabular}
\end{ruledtabular}
  \caption{Maximal rank of $\rA(t)$ for different configurations.}
         \label{tab:rank}
\end{table}
\endgroup

\section{Dual-unitary gates}

We start with quantum circuits where the two-qubit gate $U$ is of the so-called dual-unitary type~\cite{brunoprl19},

\begin{eqnarray}
  \label{eq:Wxxz}
  U_{k,k+1}&=& V_k V_{k+1} W_{k,k+1}, \\
W_{k,k+1}&=& \exp{\left(-\ii \frac{\pi}{4}( \sx_k\sx_{k+1}+\sy_k \sy_{k+1}+\az \sz_k \sz_{k+1})\right)},\nonumber \\
V_k&=&\exp{\left(-\ii (\cos{\varphi}\,\sx_k+\sin{\varphi}\, \sz_k)  \right)},\quad \varphi=0.6. \nonumber 
%&&h_x=\cos(0.6)\approx 0.82,\quad h_z=\sin(0.6)\approx 0.56.\nonumber 
\end{eqnarray}
The reason to use dual-unitary gate $W_{k,k+1}$ is that in random circuits, where $V_k$ are random gates, such evolution has been found to result in the largest change in purity relaxation rate $r$~\cite{prx21}, meaning that one could observe the two-step relaxation clearly already in smaller systems. Because our numerics is limited to relatively small system sizes with $n \le 32$ qubits it makes sense to start with a setting where one expects the effect to be large.

It is also worth mentioning that any two-qubit gate $U$ can be written in the canonical form~\cite{canonical} as $U_{k,k+1}=\tilde{V}_k \tilde{V}_{k+1} W_{k,k+1} V'_k V'_{k+1}$, where the two-qubit gate $W$ only has XYZ-like 2-body terms,
\begin{equation}
 W_{k,k+1}= {\rm e}^{-\ii \frac{\pi}{4}(\ax \sx_k\sx_{k+1}+\ay \sy_k \sy_{k+1}+\az \sz_k \sz_{k+1})}, 
\label{eq:canonical}
\end{equation}
parameterized by three $\ax,\ay,\az \in [0,1]$. Dual-unitary choice $\ax=\ay=1$  results in the fastest bipartite entanglement generation (i.e., decay of purity) for half-half bipartition~\cite{prx21} with Haar random single-qubit $V_k$, as well as for non-random gates and fixed size of subsystem A~\cite{bruno19,lorenzo20,bruno23}. We will see that this carries over also to the situation studied here, that is to half-half bipartition and non-random gates.

\begin{figure}[t!]
  \includegraphics[width=2.8in]{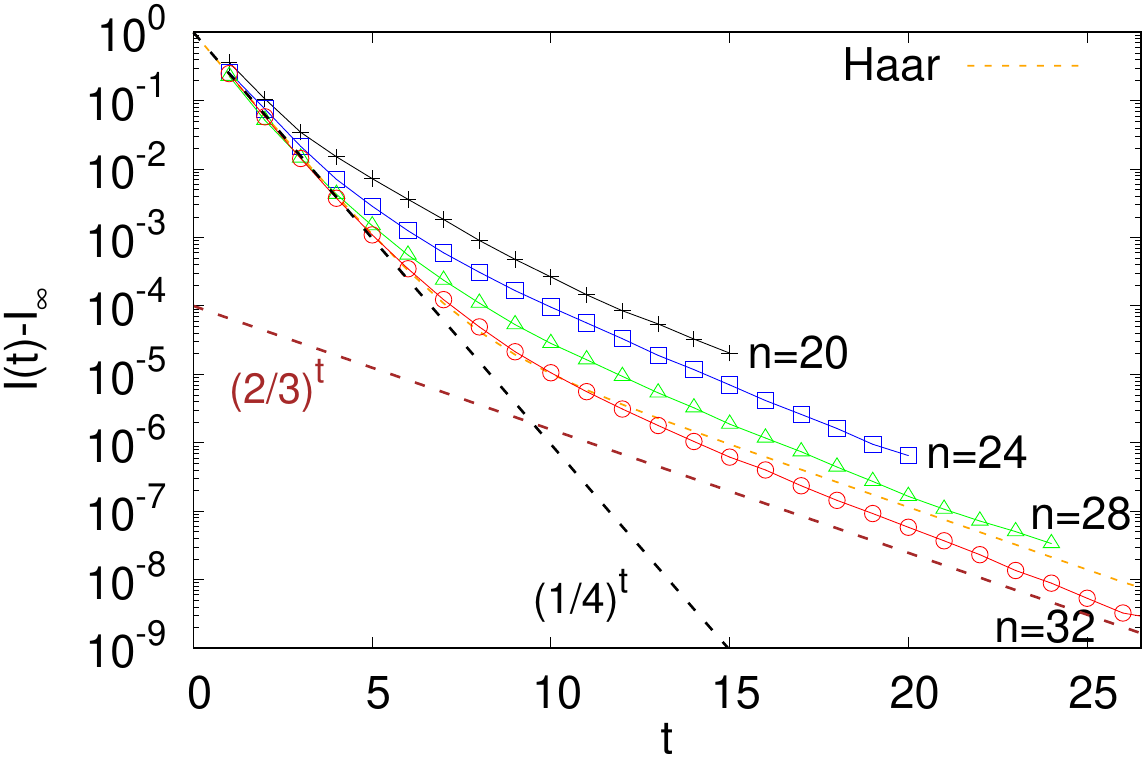}
% fajli: ~/work/rand_clif/ki/kickedxxz/xxz0.5crtapbc...
  \caption{Purity relaxation for S-PBC configuration and a dual-unitary XXZ gate with $\az=0.5$, Eq.~(\ref{eq:Wxxz}). Two dashed lines suggest behavior for $t < \tc=n/4$ (black) and for $t> \tc$ (brown). Dotted orange curve, \new{visible for $t>15$ between $n=28$ and $n=32$ points,} shows for comparison decay in the case of a random circuit ($n=32$) where $V_k$ are independent Haar random single qubit gates. All is for a single product initial state.
  }
  \label{fig:Spbc}
\end{figure}
In Fig.~\ref{fig:Spbc} we show results of numerical simulation showing a single run of $I(t)-I_\infty$ (no averaging over initial states). We can see that with increasing system size $n$ there is a sharp change in the decay slope around $\tc \approx n/4$. For a random circuit with the same two-qubit gate $W$ it was found that the asymptotic slope is given by the 2nd largest eigenvalue of the transfer matrix which is equal to $(2-\cos(\pi \az))/3=2/3$~\cite{otoc22}. The asymptotic decay in the phase II. that starts after $\tc$ seems to be rather close to this random circuit result also for our non-random Floquet system. Purity decay in the phase I. is on the other hand for large $n$ given by $I(t)=(1/4)^2$, in agreement with the maximal possible decay rate for a given rank of $\rA$ (Table~\ref{tab:rank}), and again also with the result for random circuits (Fig.10 in Ref.\cite{prx21}).

The transition time $\tc$ (for finite $n$ it is a crossover) between relaxation phases I. and II. is $\tc=n/4$ and coincides with a time when the rank of $\rA(t)$ becomes full, i.e., $2^{2\tc+1}=2^{n/2}$. Let us have a closer look at the whole spectrum $\lambda_j$ of eigenvalues of the reduced density matrix $\rA(t)$.
\begin{figure}[thb!]
  \includegraphics[width=2.6in]{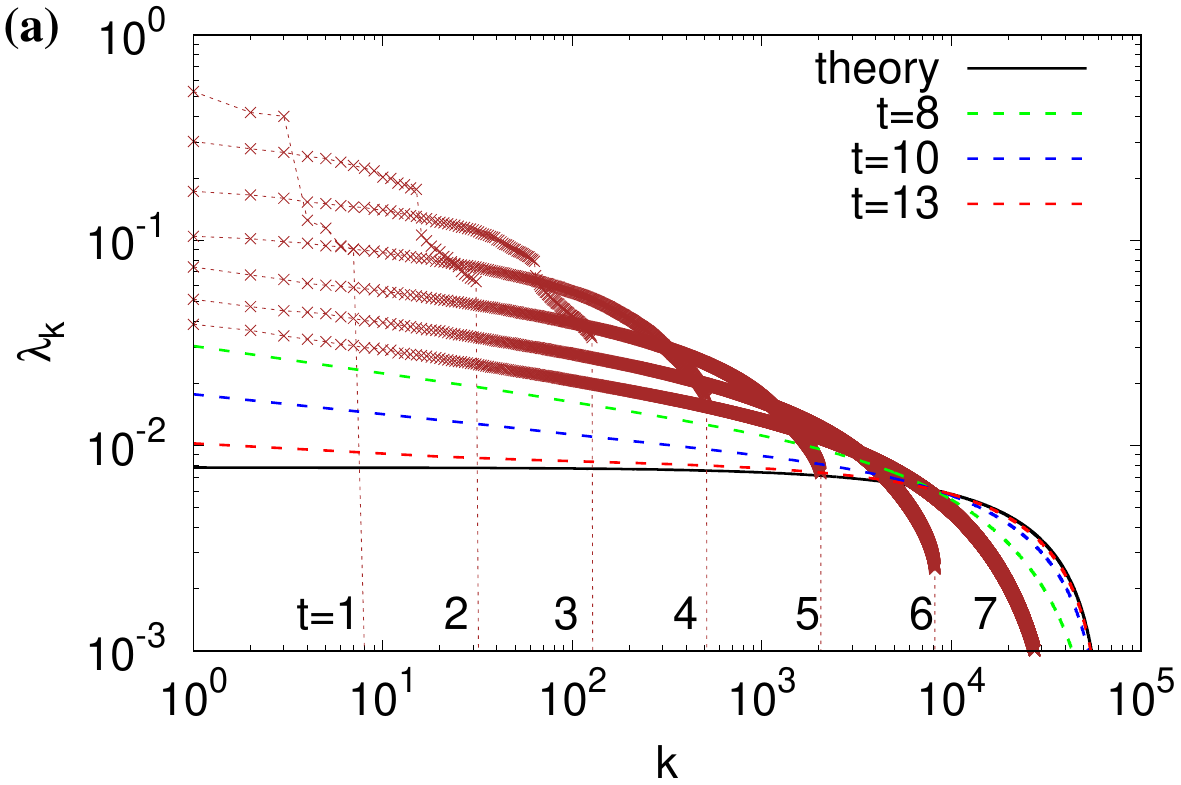}\\
  \includegraphics[width=2.6in]{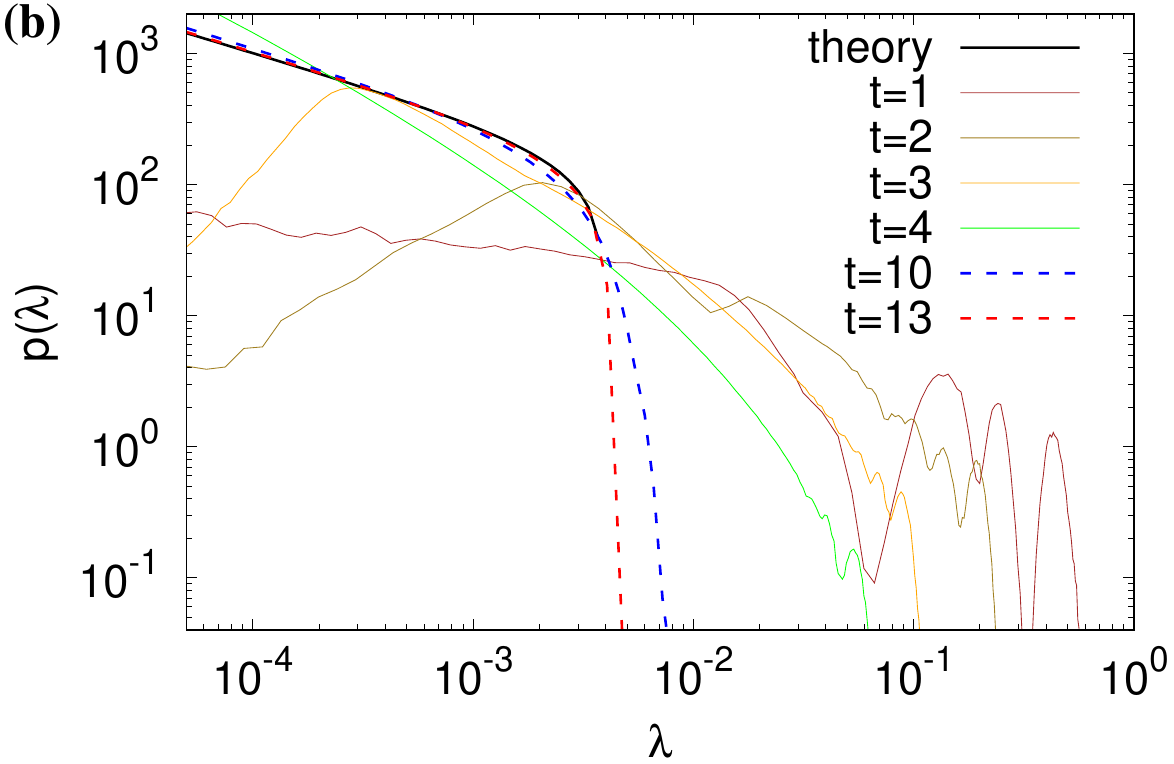}
  \caption{Eigenvalues $\lambda_k$ of $\rA$ for S-PBC configuration with $\az=0.5$ (\ref{eq:Wxxz}). (a) Evolution of the ordered eigenvalues with time. Brown crosses are at times $t<8$ when the rank is not yet full, dashed curves (green, blue, red) at full-rank times. Full black curve is theoretical random state spectrum $\lambda_k(\infty)$ (\ref{eq:lamth}) reached at $t=\infty$. (b) Eigenvalue distribution $p(\lambda)$; black full curve is the Mar\v cenko-Pastur distribution (\ref{eq:MP}). (a) is for a single initial product state with $n=32$, (b) is average over $10^4$ initial product states with $n=20$.}
  \label{fig:spec}
\end{figure}
In Fig.~\ref{fig:spec}(a) we can see how the rank increases by a factor of $4$ for every unit of time, and becomes full at $\tc=8$ for $n=32$ qubits. We can compare the spectrum to that of a random state. Density of eigenvalues $p(\lambda)$ is in the TDL the well known Mar\v cenko-Pastur distribution~\cite{MP},
\begin{equation}
  p(x)=\frac{1}{2\pi}\sqrt{(4-x)/x},\qquad x=N_{\rm A}\lambda.
  \label{eq:MP}
\end{equation}
In Fig.~\ref{fig:spec}(b) we plot the eigenvalue distribution for a couple of times. We can see that the distribution becomes close to that of random states only for $t> \tc$, that is in the phase II.. Before that $p(\lambda)$ consists of a couple of separated eigenvalues and a non-Mar\v cenko-Pastur bulk (e.g., divergence for small $\lambda$ is stronger than $1/\sqrt{\lambda}$ characteristic for Mar\v cenko-Pastur). The number of eigenvalues separated from the bulk varies with time (and circuit configuration used), e.g., for the shown S-PBC case there are 3 separated $\lambda_k$ at $t=1$ and only one at $t=4$ \new{(oscillations visible at the right edge of densities in Fig.~\ref{fig:spec}(b))}. It is also instructive to look at the average size of $k$-th largest eigenvalue (Fig.~\ref{fig:spec}(a)) and compare it to those of random states. For random states a simple result following from $p(\lambda)$ is that~\cite{ran_vec} the average $k$-th largest eigenvalue is
\begin{equation}
  \lambda_k(\infty)=\frac{4}{N_{\rm A}}\cos^2{\varphi_k},\quad  \frac{(k-\frac{1}{2})\pi}{2N_{\rm A}}=\varphi_k-\frac{1}{2}\sin{(2\varphi_k)},
  \label{eq:lamth}
\end{equation}
where the average $\lambda_k$ is expressed implicitly in terms of $\varphi_k$, where $k=1,\ldots,N_{\rm A}$. Purity, or in general $p$-th order purity $I^{(p)}={\rm tr}{\rA^p}$, can then be written as~\cite{foot2}
\begin{equation}
  I^{(p)}(t)-I^{(p)}(\infty)=\sum_k \lambda_k^p(t)-\lambda_k^p(\infty).
  \label{eq:Ip}
\end{equation}
In Fig.~\ref{fig:spec}(a) we can see how the spectrum is approaching the infinite time shape (\ref{eq:lamth}), showing together with purity data in Fig.~\ref{fig:Spbc} that the II. phase (that is for $n=32$ fully in place at about $t \approx 13$) coincides with the spectrum of $\rA$ getting very close to that of a random state.
\begin{figure*}[hbt!]
  \centerline{
    \includegraphics[width=0.3\textwidth]{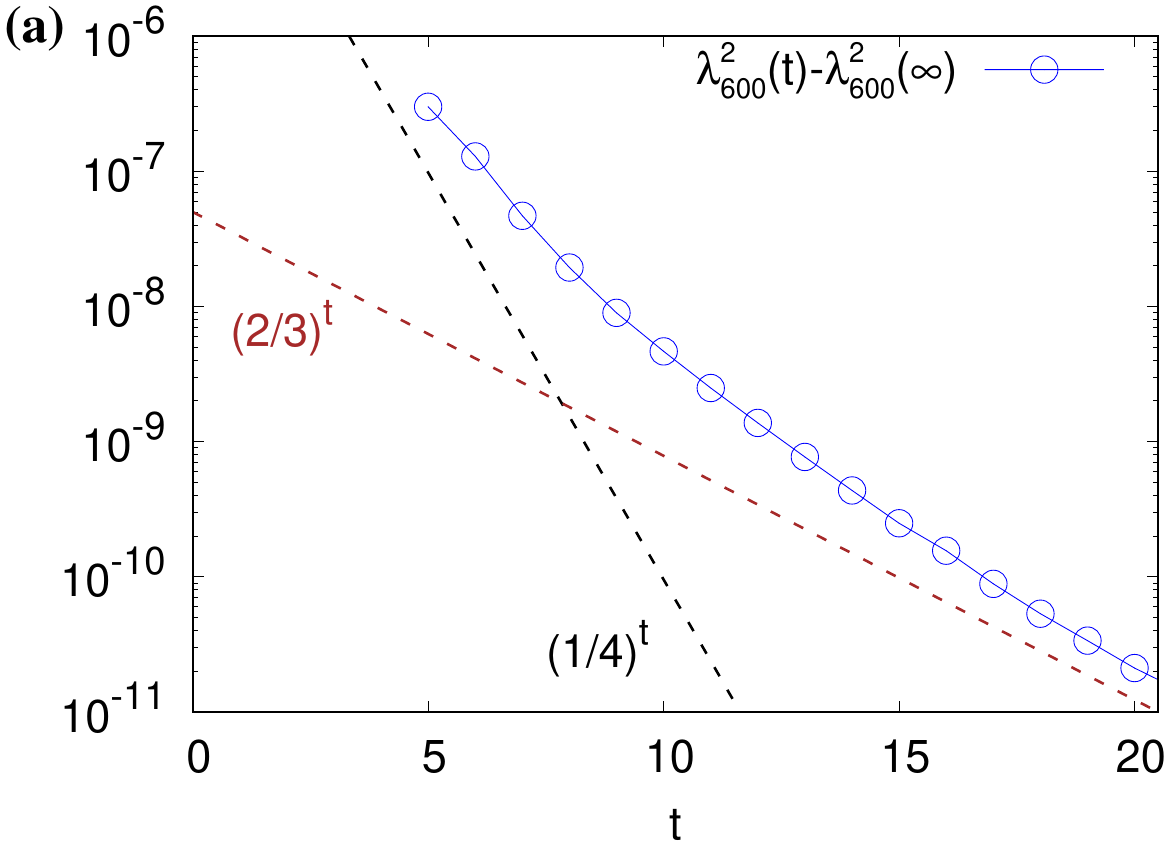}
    \includegraphics[width=0.3\textwidth]{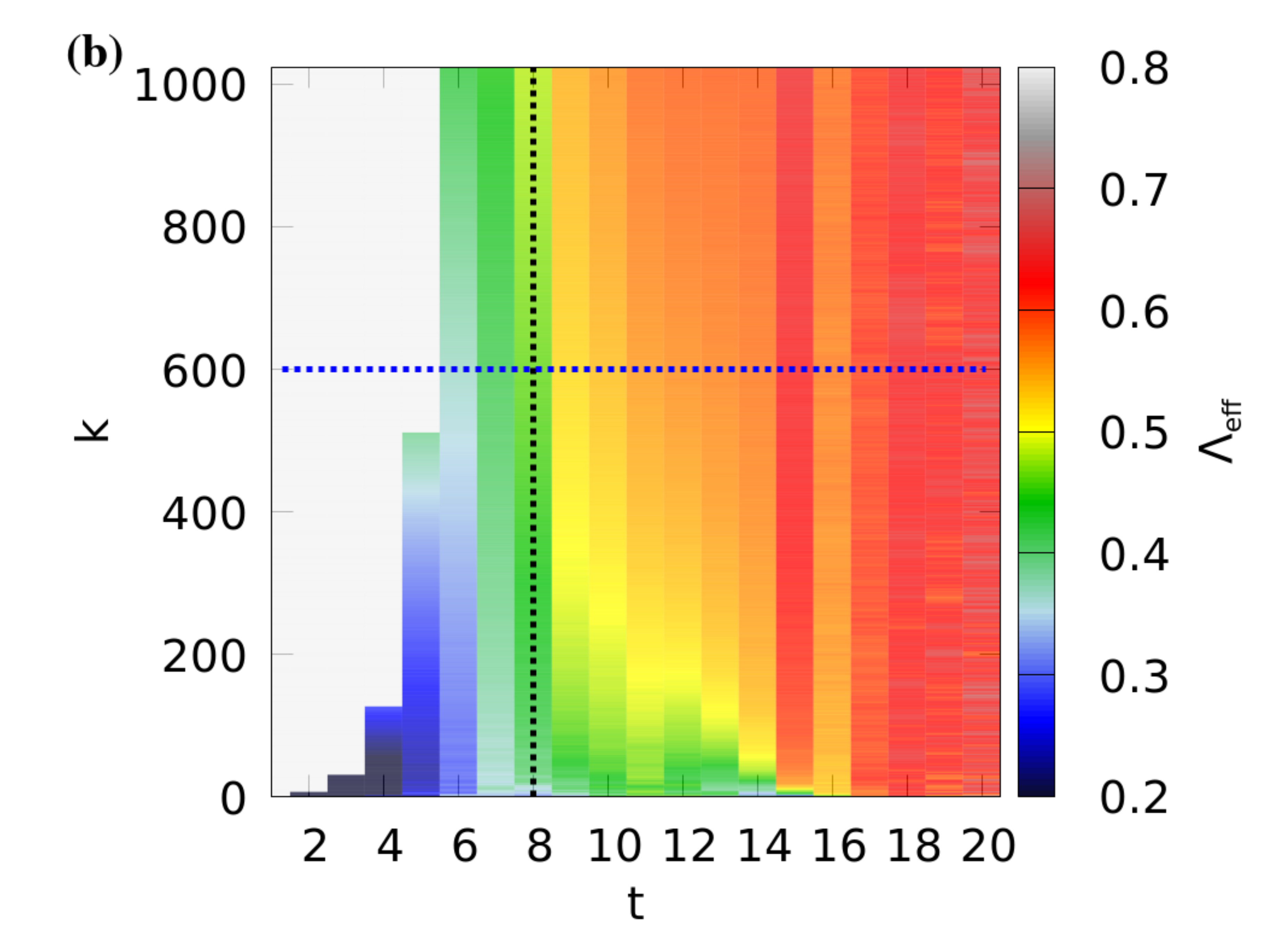}
    \includegraphics[width=0.3\textwidth]{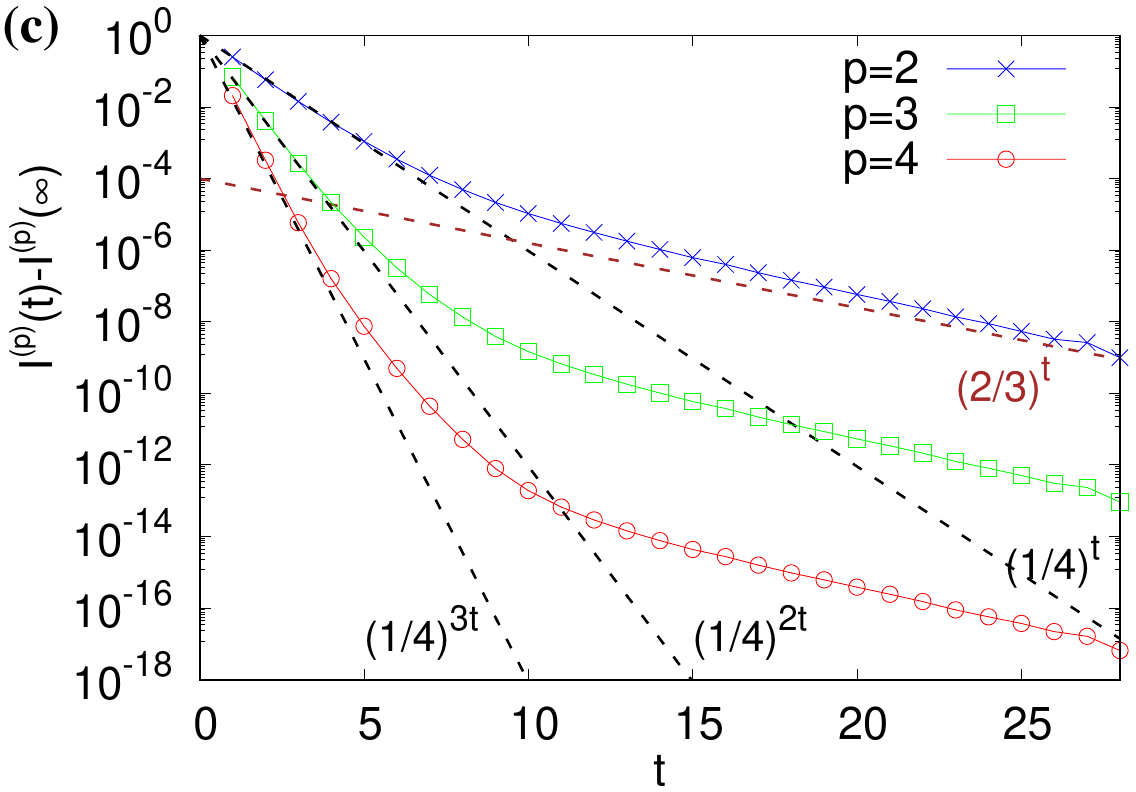}
  }
  \caption{S-PBC configuration with the XXZ gate and $\az=0.5$ (\ref{eq:Wxxz}), for one product initial state on $32$ qubits. (a) Relaxation of the $600$-th largest eigenvalue. (b) The local decay rate $\Lambda_{\rm eff}$ defined via $\lambda_k^2(t)-\lambda_k^2(\infty) \sim \Lambda_{\rm eff}^t$. Vertical dotted black line marks $\tc$, while the horizontal blue line denotes the eigenvalue shown in (a). (c) Relaxation of higher order purities $I^{(p)}(t)$ (\ref{eq:Ip}).}
  \label{fig:Ip}
\end{figure*}

Focusing more closely on individual eigenvalues we can see in Fig.~\ref{fig:Ip}(a,b) that the change in the slope visible in purity (Fig.~\ref{fig:Spbc}) is reflected also in the corresponding change in the behavior of individual eigenvalues. In Fig.~\ref{fig:Ip}(b) we see a marked change in the logarithm of the decay rate from a value close to $1/4$ for $t<\tc$ to a value closer to $2/3$ for $t>\tc$. Considering that in all figures so far (except Fig.~\ref{fig:spec}(b)) we were showing a single realization (i.e., no averaging), it is clear that the same 2-phase relaxation would be observed also in higher order purities or the logarithms of $I^{(p)}$ called R\' eny entropies. Fig.~\ref{fig:Ip}(c) shows that $I^{(p)}$ also approach their infinite time values $I^{(3)}(\infty)=(5N_{\rm A}^2+1)/((N_{\rm A}^2+1)(N_{\rm A}^2+2))$ and $I^{(4)}(\infty)=(14N_{\rm A}^3+10N_{\rm A})/((N_{\rm A}^2+1)(N_{\rm A}^2+2)(N_{\rm A}^2+3))$~\cite{karol01} in two phases with a kink at $\tc\approx n/4$. What is more, it looks that for $t<\tc$ higher order purities decay in the TDL as simple powers, $I^{(p)} \sim (1/4)^{(p-1)t}$. Such decay is equal to the maximal possible decay rate of $I^{(p)}$ compatible with a finite rank of $\rA$ (Table~\ref{tab:rank}), and would be realized if the spectrum would be flat, $\lambda_k(t)=1/({\rm rank})$. While the spectrum (Fig.~\ref{fig:spec}(a)) is not exactly flat, it seems that the circuit is nevertheless scrambling enough to result is such leading order behavior in the TDL. We also note that fluctuations of purity between different choices of random product initial states are small. For instance, for data in Fig.~\ref{fig:Spbc} one has $\sigma(I)<I$ for all $t\le n$. Therefore, unless one is interested in times longer than $\approx 4\tc$ one initial state is enough to observe the phenomenon\new{, and as a consequence all R\' eny entropies behave the same}. Similar independence on its order $p$ has been obtained in the TDL also for R\' eny entropies in the case of a self-dual kicked Ising model for a bipartition with a finite size subsystem A~\cite{bruno19}, see also Ref.~\cite{lorenzo20}.
\begin{figure*}[bth!]
  \centerline{
    \includegraphics[width=0.3\textwidth]{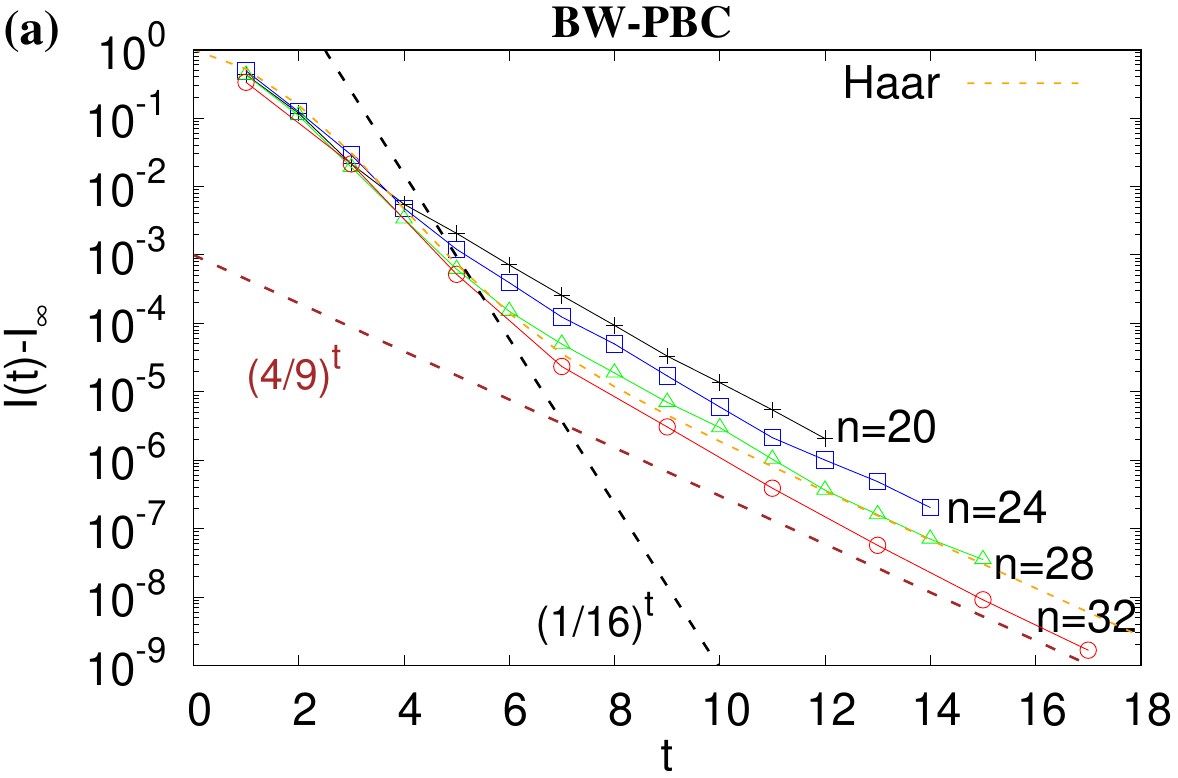}
    \includegraphics[width=0.3\textwidth]{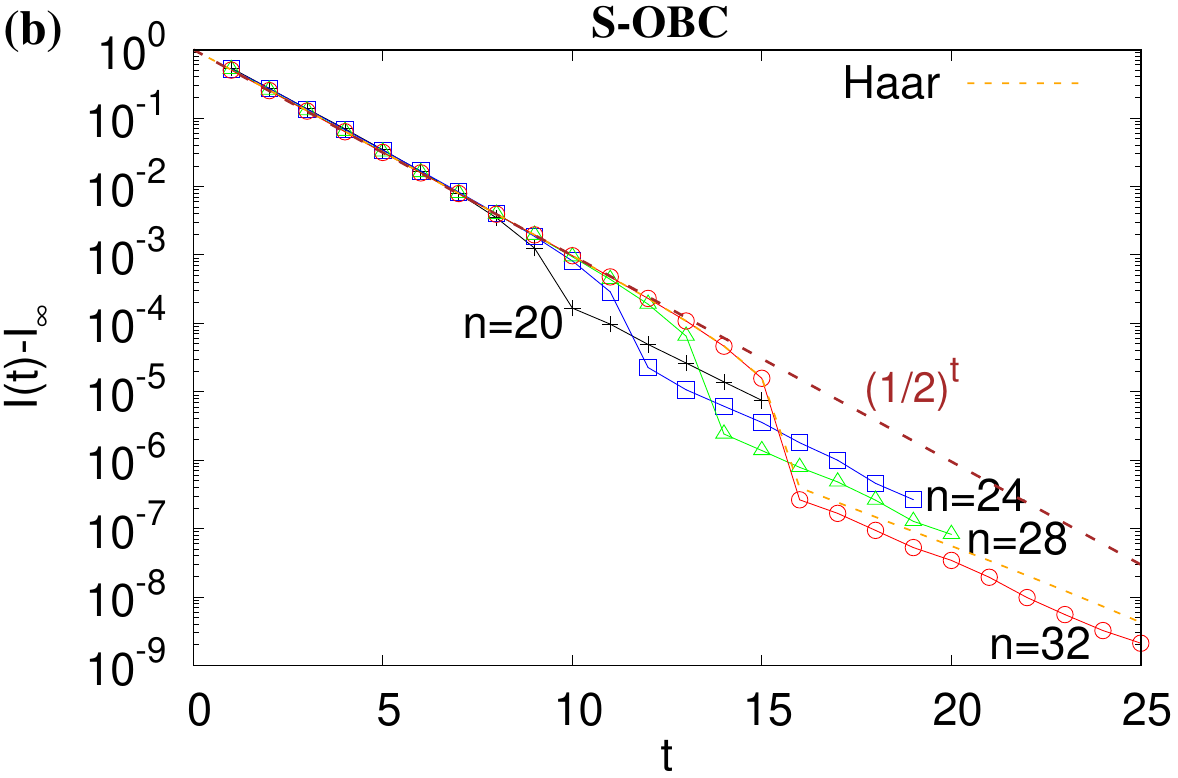}
    \includegraphics[width=0.3\textwidth]{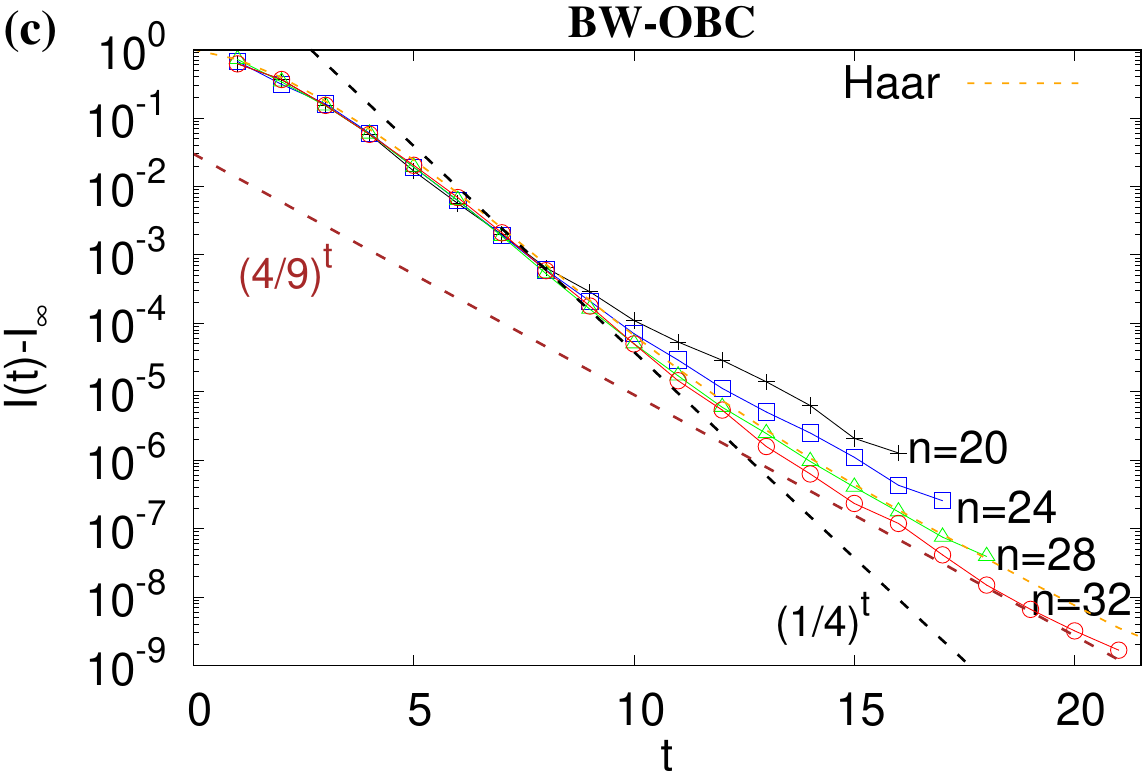}
  }
  \caption{Purity relaxation for different protocols with the dual-unitary XXZ gate with $\az=0.5$ (\ref{eq:Wxxz}), all for a single product initial condition. Protocols BW-PBC (a), S-OBC (b), and BW-OBC (c) are shown. Orange dotted curve is data for a circuit with Haar random single qubit gates and $n=32$.}
  \label{fig:kicked}
\end{figure*}

So far we have shown data for the S-PBC protocol. In Fig.~\ref{fig:kicked} we show purity relaxation for other S and BW protocols and boundary conditions. As one can see relaxation is in all cases rather similar to the one for a random circuit in which all single qubit gates are independent Haar random gates, see Ref.~\cite{prx21}. The transition time $\tc$ depends on the configuration and is in line with the growth of the rank of $\rA$, that is $\tc \approx n/8$ for BW-PBC, $\tc \approx n/2$ for S-OBC, and $\tc \approx n/4$ for BW-OBC. An interesting feature can be seen for S-OBC where there is a discontinuous jump at $\tc$ rather than a change in the slope, similar as in random circuits~\cite{prx21} \new{with XXZ two qubit gates and S-OBC configuration}. Such jump in S-OBC is likely specific to dual-unitary circuits and their special propagation properties, for instance, infinite-temperature 2-point correlations are nonzero only on the lightcone boundary~\cite{brunoprl19}. Another specific point to S-OBC is also that for smaller $\az$, e.g $\az=0.2$ (data not shown), fluctuations would become relevant, \new{that is $\sigma(I) \approx I$}, already at $\tc$.

With a rather limited range of system sizes $n$ for which we have data available we are not able to conclusively demonstrate that the transition at $\tc$ is sharp in the TDL, i.e., that the transition is discontinuous in the scaled time $t/n$. However, we note that in a circuit with Haar random single-qubit gates, where a Markovian description allows for simulation of larger $n$, the transition is sharp (see Fig.10(b) in Ref.~\cite{prx21}). The transition in fact appears to be sharp in the TDL in all~\cite{fut} random circuits with a fixed 2-qubit gate and Haar random single-qubit gates~\cite{prx21}, see also our Fig.~\ref{fig:xxzSobc} where the transition is quite sharp already at $n=32$.

Final comment we want to make in this section is whether the transition in the spectrum of $\rA$ at $\tc$ can be described by any of the studied random matrix/state ensembles? Namely, transitions have been observed before in specific ensembles. One case is if the evolution is given by a random Hamiltonian~\cite{vinayak,brandao} in which case the distribution of the largest eigenvalue exhibits a transitions, while the spectrum can be described at short times by a Mar\v cenko-Pastur-like bulk plus one separated eigenvalue~\cite{vinayak}. The ensemble that can describe such behavior is the correlated Wishart ensemble (the uncorrelated Wishart ensemble describes random states). Not surprisingly, in our case behavior is different; after all the locality of our propagator $U(1)$ and non-commutativity of individual gates is definitely important and can not be modeled by $\exp{(-\ii Ht)}$ with random $H$. This is clearly reflected also in the spectrum where the bulk is non-random and there can be several separated eigenvalues, not just one. Transitions have been observed also in a finite-''temperature'' random states ensemble~\cite{anto10}, or, related, in the large deviation properties of purity distribution (e.g., distribution of purity exhibits three different phases)~\cite{Nadal11}, but they also have different properties than our case. We also did not find a connection between $\tc$ and a maximum in the bandwidth of the spectrum, as observed in Ref.~\cite{jed} (for some circuits the bandwidth does exhibit a sharp maximum, for other it does not).

\section{General gates}

In previous section we used a two-qubit $U$ that had a form of a kicked system (\ref{eq:Wxxz}), that is, it was a product of single-qubit gates $V_k$ and a two-qubit gate $W_{k,k+1}$. To make a generic non dual unitary gate we shall simply write the two-qubit $U_{k,k+1}$ in an exponential form,
\begin{eqnarray}
  \label{eq:2qxxz}
U_{k,k+1}&=&\exp{\left(-\ii H\right)}\\
H&=&\frac{\pi}{4} (\sx_k\sx_{k+1}+\sy_k \sy_{k+1}+\az \sz_k \sz_{k+1})+\nonumber \\
&&+h_x(\sx_k+\sx_{k+1})+h_z (\sz_k+\sz_{k+1}),\nonumber
\end{eqnarray}
with the same $h_x=\cos{\varphi}$, $h_z=\sin{\varphi}$ and $\varphi=0.6$ as before (\ref{eq:Wxxz}). Note that such $U$ is not of the dual-unitary form; writing it in the canonical form (\ref{eq:canonical}) the parameters would e.g. be $(\ax,\ay,\az)\approx (1.00,0.90,0.60)$ for $\az=0.5$, and $(1.00,0.84,0.37)$ for $\az=0.2$. The reason to still use the ``maximal'' prefactor $\pi/4$ in Eq.~(\ref{eq:2qxxz}) is to have fast purity decay and therefore hopefully large effect that is possible to clearly observe already in systems with $n \le 32$.
\begin{figure}[t!]
  \includegraphics[width=2.8in]{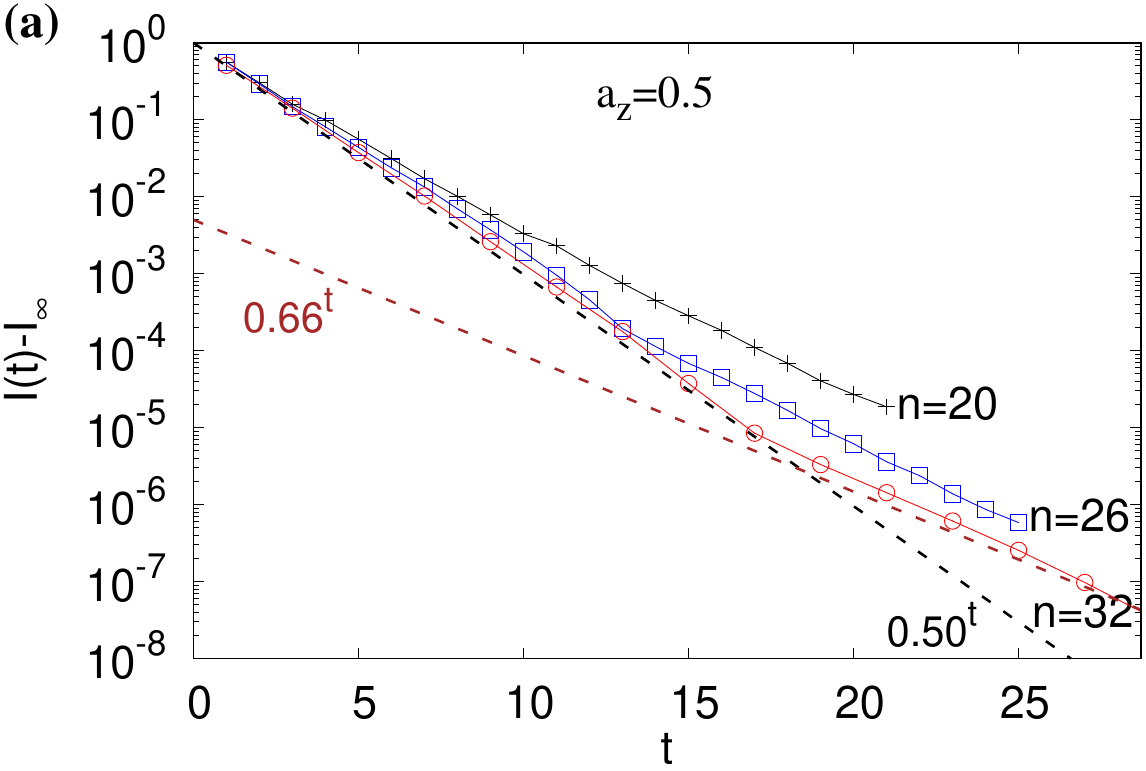}
  \includegraphics[width=2.8in]{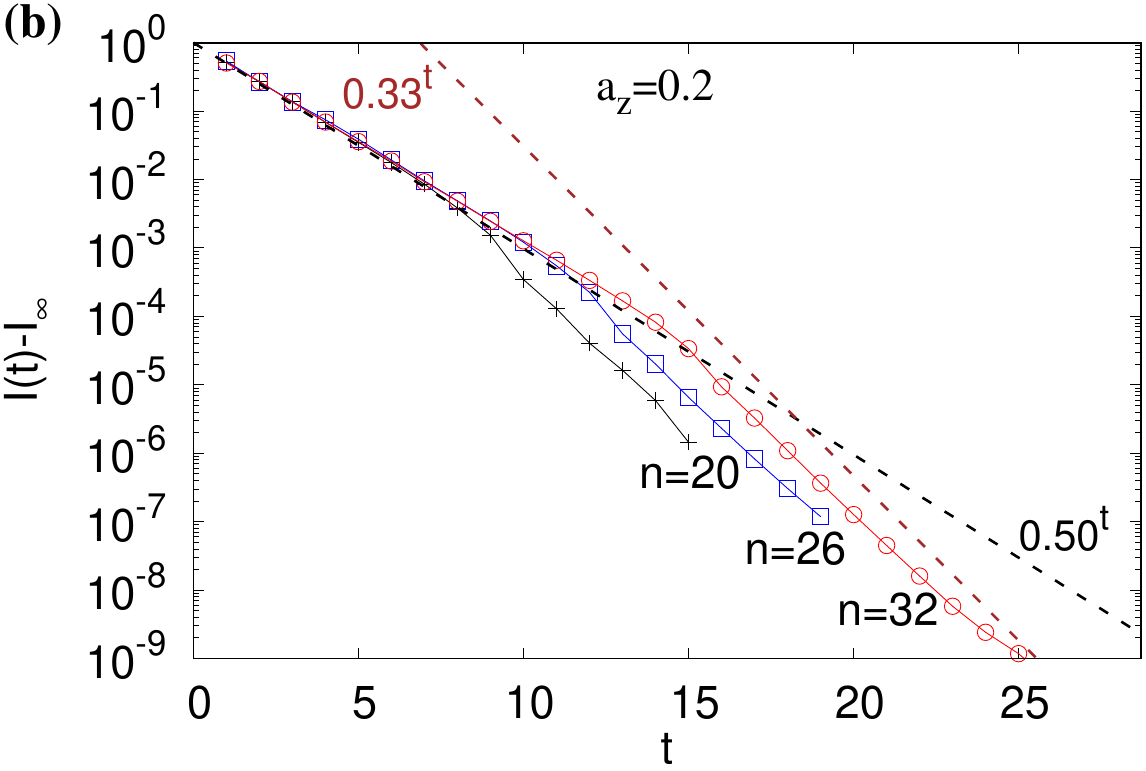}
  \caption{Purity relaxation in a circuit with an XXZ gate (Eq.~\ref{eq:2qxxz}) in the S configuration with OBC (one product initial state). (a) For $\az=0.5$ one gets initially faster decay, while in (b) for $\az=0.2$ one initially has slower decay.}
  \label{fig:xxzSobc}
\end{figure}
In Fig.~\ref{fig:xxzSobc} we show purity relaxation for the S circuit with OBC and $\az=0.5$ as well as $\az=0.2$. In both cases there is again a change in the slope at $\tc$ which in this case scales as $\tc \approx n/2$, in line with the finite rank (Table~\ref{tab:rank}): the transition happens at $\tc$ when the spectrum of $\rA$ becomes close to that of random states and $I(t)$ would be close to its saturation value $I_\infty$. The initial decay rate upto $\tc$ is equal to the maximal possible rate for the OBC staircase configuration (Table~\ref{tab:rank}). While for $\az=0.5$ the subsequent phase II. decay is slower, at $\az=0.2$ it is interestingly faster than the initial rate in the phase I., similar to the so-called phantom decay~\cite{prx21,skin22} in random circuits in which the initial decay is slower than the gap of the Markovian transfer matrix would suggest. 
\begin{figure}[t!]
  \includegraphics[width=2.8in]{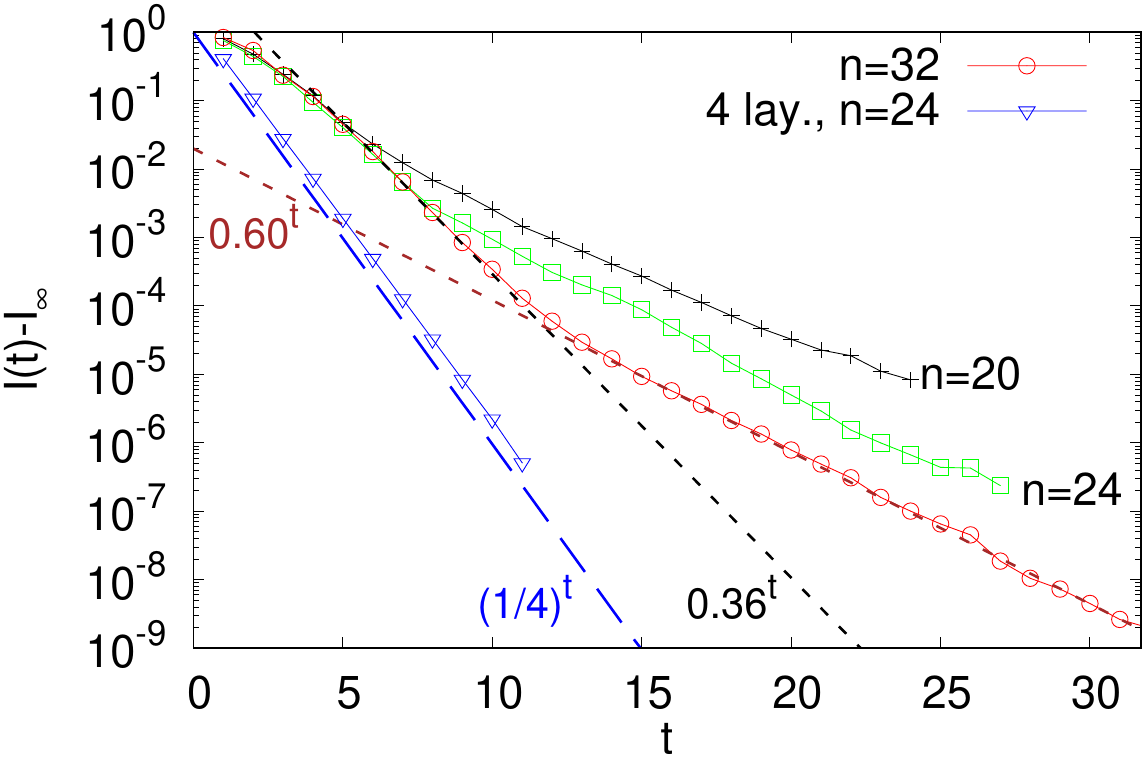}
% fajli: ~/work/rand_clif/ki/kickedxxz/2qxxz0.5-BWobc...
  \caption{BW circuit with OBC and the XXZ gate (\ref{eq:2qxxz}) with $\az=0.5$. Initial faster decay in phase I. transitions at $\tc$ into slower decay in phase II.. Blue triangles show for comparison purity decay for a circuit that has per unit of time 4 BW layers in subsystem A, as well as in B, and a single gate coupling A and B.}
  \label{fig:xxzBWobc}
\end{figure}
In Fig.~\ref{fig:xxzBWobc} we show results for the BW circuit with OBC. Compared to the dual-unitary gate in Fig.~\ref{fig:kicked}(c) the change in the rate between phases I. and II. is here even larger. The decay in the phase I. here seem to be slower than the maximal possible one ($1/4^t$), though it could be that the largest $n=32$ we show is still too small to fully reach the TDL.

One natural question is what is the microscopic origin of differing relaxation rates before and after $\tc$? Considering that $\rA$ is not yet full rank for $t<\tc$ one could speculate that the crucial difference is precisely that -- a finite rank of the reduced density matrix. Using a Coulomb gas picture for the evolution of eigenvalues it is suggestive to say that in the phase I. the eigenvalues ``expand'' into a ``vacuum'' (the number of eigenvalues is smaller than $2^{n/2}$), \new{and therefore evolve differently than in the phase II. where all eigenvalues are already nonzero} (existing full-rank eigenvalues serve as a ``bath''). However, this can not be a full story: we illustrate that with a circuit in which one again has the same finite rank $\rA$ but without strong locality of only $\sim n/2$ gates acting on subsystems A and B, and which does not have any change in the slope at $\tc$ (blue triangles in Fig.~\ref{fig:xxzBWobc}). Let us write BW-OBC circuit as $U(1)=U_{\rm B} U_{n/2,n/2+1} U_{\rm A}$, where $U_{\rm A/B}$ denotes all the gates acting only on A, or B. Taking now a circuit with a single-step propagator $U(1)=U_{\rm B}^4 U_{n/2,n/2+1} U_{\rm A}^4$ (shown with blue triangles in Fig.~\ref{fig:xxzBWobc}), i.e., 4 brick-wall layers in A and B, one has exactly the same rank of $\rA$ at time $t$ as for our standard BW-OBC. However, due to 4 layers in A dynamics there looses its strong local character being present in the case with a single BW layer. We can see that this causes purity to decay in the fastest possible way as $(1/4)^t$ and seemingly without any change in the slope. This shows that locality and non-commutativity of gates \new{seems to be} crucial for the two-step relaxation. The importance of non-commutativity is nicely seen in random circuits whose average dynamics can be described by a Markovian process and where the important effects of non-Hermiticity come due to non-commutativity of individual gates~\cite{prx21,skin22}. \new{While Hamiltonian dynamics requires an independent study, based on what we learned we might speculate} that the effect will go away in the Hamiltonian limit of two-qubit gates in which one would decrease the prefactor in gates from $\pi/4$ towards $0$, i.e., replace $U(1)$ by $\exp{(-\ii H)}$ with some local $H$. We have verified though (data not shown) that the two-step relaxation is not limited to the prefactor \new{(time) in two-qubit gates} being exactly $\pi/4$ (for instance, using $0.7 \pi/4$ in Eq.(\ref{eq:Wxxz}) with S-PBC, or $0.8 \pi/4$ in Eq.(\ref{eq:2qxxz}) with S-OBC, still results in a two-step relaxation).

\subsection{OTOC functions}

\begin{figure}[b!]
  \includegraphics[width=2.8in]{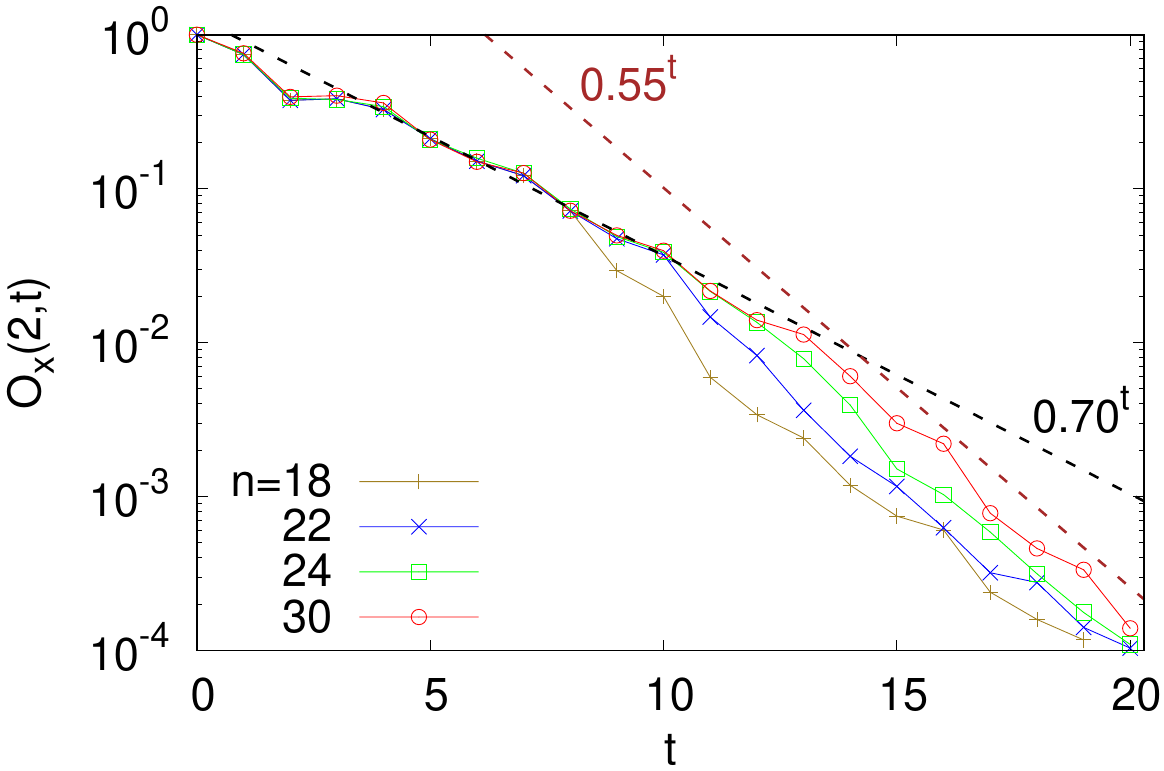}
  \caption{OTOCs decay for the BW circuit with PBC and XXZ gate with $\az=0.4$, Eq.~(\ref{eq:2qxxz}). Here averaging over initial random product states is performed (1000 for $n=18$, 4 for $n=30$).}
  \label{fig:otoc}
\end{figure}

So far we have studied purity decay. Purity is quadratic in $U$ and $U^\dagger$, similarly as is the out-of-time-ordered (OTOC) correlation function,
\begin{equation}
  O_{\alpha}(j,t)=\ave{ \sigma_j^{\rm z}(t)\sigma_1^\alpha \sigma_j^{\rm z}(t)\sigma_1^\alpha },\quad \sigma_j^{\rm z}(t)=U^\dagger(t) \sigma_j^{\rm z} U(t),
  \label{eq:otoc}
\end{equation}
where the averaging is done over random product initial states. In random circuits the two-step relaxation has been observed also in the OTOC function decay~\cite{otoc22}. We shall now demonstrate that the same happens also in non-random Floquet circuits \new{-- one can have a two-step relaxation also in OTOCs.}

In Fig.~\ref{fig:otoc} we can see that the OTOC function (\ref{eq:otoc}) evaluated for $\sigma_2^{\rm x}$ exhibits a two-step relaxation (similar result would be obtained also for other sites, or Pauli matrices). Upto time $\tc$ proportional to $n$ the decay is slower, after which it goes into faster relaxation.

\section{Conclusion}

We have shown numerically that in a number of quantum circuits in which the Floquet propagator is composed of the same nearest-neighbor gates in e.g. brick-wall or staircase configuration relaxation of purity to its long-time random state value proceeds in two phases. In the first phase the rank of the reduced density matrix is deficient (less than full), while in the 2nd phase in which relaxation proceeds with a different rate, $\rA$ has a full rank. The critical time between the two phase is proportional to the number of qubits. In the 1st phase the spectrum of $\rA$ is still far away from that of random states, while in the 2nd phase $\rA$ is already close to being random. Because the critical transition time is proportional to $n$ one can also say that it happens when the system realizes it is finite. In this sense it actually looks rather natural that there should be two different relaxation times, one before correlations propagate to the boundary, and one after that.

Relaxation proceeding in two steps is not limited to purity, we have observed it also in out-of-time ordered correlation functions. While both quantities are quadratic in $\rho(t)$, the effect is likely not limited to such objects. That this is the case can be argued based on self-averaging in large systems, and \new{furthermore has been explicitly shown for higher order purities (i.e., R\' eny entropies)}. It would be interesting to get a microscopic physical picture behind the effect. \new{To that end an appropriate statistical ensemble (of not fully random) states that would exhibit the same transition would be useful.

 The results presented show that the two-step relaxation is not limited to just random circuits where the average dynamics is Markovian, as observed previously. While the effect in random circuits can be traced to an explicit non-Hermitian transfer matrix and the associated localized eigenvectors, here no such analytical simplification is possible, nevertheless, the effect is still present.} 

We would like to acknowledge support by Grants J1-4385 and No.~P1-0402 from the Slovenian Research Agency.


\begin{thebibliography}{99}

\bibitem{landau} L.~D.~Landau and E.~M.~Lifshitz, {\em Statistical physics} (Pergamon, 1969).
  % p384 okoli eq.(121.12)

\bibitem{Haake} F.~Haake, {\em Quantum signatures of chaos} (Springer, 2010).

\bibitem{felix} F.~Borgonovi, F.~M.~Izrailev, L.~F.~Santos, and V.~G.~Zelevinsky, \tit{Quantum chaos and thermalization in isolated systems of interacting particles} Phys.~Rep. {\bf 626}, 1 (2016).

\bibitem{neumann} J.~{von Neumann}, \tit{Beweis des Ergodensatzes und des H-Theorems in der neuen Mechanik} Z.~Phys. {\bf 57}, 30 (1929); J.~{von Neumann}, \tit{Proof of the ergodic theorem and the H-theorem in quantum mechanics} Eur.~Phys.~J.~H {\bf 35}, 201 (2010).
  
\bibitem{Hayden} P.~Hayden, D.~W.~Leung, and A.~Winter, \tit{Aspects of generic entanglement} Commun.~Math.~Phys. {\bf 265}, 95 (2006).

\bibitem{joel} S.~Goldstein, J.~L.~Lebowitz, C.~Mastrodonato, R.~Tumulka, and N.~Zanghi, \tit{Normal typicality and von {N}eumann’s quantum ergodic theorem} Proc.~R.~Soc. {\bf 466}, 3203 (2010).

\bibitem{eth} M.~Srednicki, \tit{Chaos and quantum thermalization} Phys.~Rev.~E {\bf 50}, 888 (1994).

\bibitem{anatoli} A.~Polkovnikov, K.~Sengupta, A.~Silva, and M.~Vengalattore, \tit{Colloquium: Nonequilibrium dynamics of closed interacting quantum systems} Rev.~Mod.~Phys.~ {\bf 83}, 863 (2011).

\bibitem{gogolin} C.~Gogolin and J.~Eisert, \tit{Equilibration, thermalisation, and the emergence of statistical mechanics in closed quantum systems} Rep.~Prog.~Phys. {\bf 79}, 056001 (2016).
  
\bibitem{casati} G.~Casati, B.~V.~Chirikov, F.~M.~Izrailev, and J.~Ford, \tit{Stochastic behavior of a quantum pendulum under a periodic perturbation} in Lecture Notes in Physics {\bf 93}, p.334 (Springer, Berlin, 1979).

\bibitem{ulrich} U.~Schollw\" ock, \tit{The density-matrix renormalization group in the age of matrix product states} Annals of Physics {\bf 326}, 96 (2011).  


\bibitem{adam18} A.~Nahum, S.~Vijay, and J.~Haah, \tit{Operator spreading in random unitary circuits} Phys.~Rev.~X {\bf 8}, 021014 (2018).

\bibitem{vedika} V.~Khemani, A.~Vishwanath, D.~A.~Huse, \tit{Operator spreading and the emergence of dissipative hydrodynamics under unitary evolution with conservation laws} Phys.~Rev.~X {\bf 8}, 031057 (2018).

\bibitem{chalker18} A.~Chan, A.~{De Luca}, and J.~T.~Chalker, \tit{Solution of a minimal model for many-body quantum chaos} Phys.~Rev.~X {\bf 8}, 041019 (2018).
  
\bibitem{tianci} T.~Zhou and A.~Nahum, \tit{Emergent statistical mechanics of entanglement in random unitary circuits} Phys.~Rev.~B {\bf 99}, 174205 (2019).

\bibitem{markov} W.-T.~Kuo, A.~A.~Akhtar, D.~P.~Arovas, and Y.~Z.~You, \tit{Markovian entanglement dynamics under locally scrambled quantum evolution} Phys. Rev. B {\bf 101}, 224202 (2020).

\bibitem{skin22} M.~\v Znidari\v c, \tit{Solvable non-Hermitian skin effect in many-body unitary dynamics} Phys.~Rev.~Research {\bf 4}, 033041 (2022).

\bibitem{skin23} M.~\v Znidari\v c, \tit{Phantom relaxation rate of the average purity evolution in random circuits due to Jordan non-Hermitian skin effect and magic sums} Phys.~Rev.~Research {\bf 5}, 033145 (2023).
  % DOI https://doi.org/10.1103/PhysRevResearch.5.033145
  %\new{M.~\v Znidari\v c, \tit{Phantom relaxation rate due to Jordan non-Hermitian skin effect and magic sums} {\tt arXiv:2306.07876} (2023).}  

\bibitem{brunoprl19} B.~Bertini, P.~Kos, and T.~Prosen, \tit{Exact correlation functions for dual-unitary lattice models in 1 + 1 dimensions} Phys.~Rev.~Lett. {\bf 123}, 210601 (2019).

\bibitem{bruno19} B.~Bertini, P.~Kos, and T.~Prosen, \tit{Entanglement spreading in a minimal model of maximal many-body quantum chaos} Phys.~Rev.~X {\bf 9}, 021033 (2019).
  % self-dual KI in S_r za fiksen |A|

\bibitem{bruno23} \new{A.~Foligno and B.~Bertini, \tit{Growth of entanglement of generic states under dual-unitary dynamics} Phys.~Rev.~B {\bf 107}, 174311 (2023).}

\bibitem{sarang} S.~Gopalakrishnan and A.~Lamacraft, \tit{Unitary circuits of finite depth and infinite width from quantum channels} Phys.~Rev.~B {\bf 100}, 064309 (2019).

\bibitem{katja} K.~Klobas, B.~Bertini, and L.~Piroli, \tit{Exact thermalization dynamics in the "Rule 54" quantum cellular automaton} Phys.~Rev.~Lett. {\bf 126}, 160602 (2021).
  
\bibitem{lamacraft21} P.~W.~Claeys, J.~Herzog-Arbeitman, and A.~Lamacraft, \tit{Correlations and commuting transfer matrices in integrable unitary circuits} SciPost Phys. {\bf 12}, 007 (2022).  

  
\bibitem{bloch} I.~Bloch, J.~Dalibard, and W.~Zwerger, \tit{Many-body physics with ultracold gases} Rev.~Mod.~Phys. {\bf 80}, 885 (2008).  

\bibitem{google} X.~Mi et al., \tit{Information scrambling in quantum circuits} Science {\bf 374}, 6574 (2021).

\bibitem{prx21} J.~Bensa and M.~\v Znidari\v c, \tit{Fastest local entanglement scrambler, multistage thermalization, and a non-Hermitian phantom} Phys.~Rev.~X {\bf 11}, 031019 (2021).

\bibitem{otoc22} J.~Bensa and M.~\v Znidari\v c, \tit{Two-step phantom relaxation of out-of-time-ordered correlations in random circuits} Phys.~Rev.~Research {\bf 4}, 013228 (2022).

\bibitem{bipart} J.~Bensa and M.~\v Znidari\v c, \tit{Purity decay rate in random circuits with different configurations of gates} Phys.~Rev.~A {\bf 107}, 022604 (2023).
%  {\tt arXiv:2211.13565} (2022).
  
\bibitem{skin1} L.~E.~F.~{Foa Torres}, \tit{Perspective on topological states of non-Hermitian lattices} J.~Phys.~Mater. {\bf 3}, 014002 (2020).
  
\bibitem{skin2} E.~J.~Bergholtz, J.~C.~Budich, and F.~K.~Kunst, \tit{Exceptional topology of non-Hermitian systems} Rev.~Mod.~Phys. {\bf 93}, 015005 (2021).

\bibitem{trefethen} L.~N.~Trefethen and M.~Embree, {\em Spectra and pseudospectra}, (Princeton University Press, 2005).

\bibitem{song} F.~Song, S.~Yao, and Z.~Wang, \tit{Non-Hermitian skin effect and chiral damping in open quantum systems} Phys.~Rev.~Lett. {\bf 123}, 170401 (2019).
  
\bibitem{takashi20} T.~Mori and T.~Shirai, \tit{Resolving a discrepancy between Liouvillian gap and relaxation time in boundary-dissipated quantum many-body systems} Phys.~Rev.~Lett. {\bf 125}, 230604 (2020).

\bibitem{takashi21} T.~Mori, \tit{Metastability associated with many-body explosion of eigenmode expansion coefficients} Phys.~Rev.~Res. {\bf 3}, 043137 (2021).
  
\bibitem{ueda21} T.~Haga, M.~Nakagawa, R.~Hamazaki, and M.~Ueda, \tit{Liouvillian skin effect: slowing down of relaxation processes without gap closing} Phys.~Rev.~Lett. {\bf 127}, 070402 (2021).


\bibitem{lorenza21} V.~P.~Flynn, E.~Cobanera, and L.~Viola, \tit{Topology by dissipation: Majorana bosons in metastable quadratic markovian dynamics} Phys.~Rev.~Lett. {\bf 127}, 245701 (2021).

\bibitem{clerk} G.~Lee, A.~McDonald, and A.~Clerk, \tit{Anomalously large relaxation times in dissipative lattice models beyond the non-Hermitian skin effect} {\tt arXiv:2210.14212} (2022).

\bibitem{Ho23} \new{W.~W.~Ho, T.~Mori, D.~A.~Abanin, and E.~G.~D~Torre, \tit{Quantum and classical Floquet prethermalization} Annals of Physics {\bf 454}, 169297 (2023).}

  
\bibitem{karol01} K.~Zyczkowski and H.-J.~Sommers, \tit{Induced measures in the space of mixed quantum states} J.~Phys.~A {\bf 34}, 7111 (2001).
  % I_2,I_3,I_4 infty za half-half
 
  
\bibitem{nielsen} M.~A.~Nielsen and I.~L.~Chuang, {\em Quantum computation and quantum information}, (Cambridge University Press, 2000).
  

\bibitem{max} C.~H.~Bennett, A.~W.~Harrow, D.~W.~Leung, and J.~A.~Smolin, \tit{On the capacities of bipartite hamiltonians and unitary gates} IEEE Trans.~Inf.~Theory {\bf 49}, 1895 (2003).


\bibitem{canonical} A.~M.~Childs, H.~L.~Haselgrove, and M.~A.~Nielsen, \tit{Lower bounds on the complexity of simulating quantum gates} Phys.~Rev.~A {\bf 68}, 052311 (2003).

  
\bibitem{lorenzo20} L.~Piroli, B.~Bertini, J.~I.~Cirac, and T.~Prosen, \tit{Exact dynamics in dual-unitary quantum circuits} Phys.~Rev.~B {\bf 101}, 094304 (2020).
  % S_r za fiksen |A| v TDL

\bibitem{MP} V.~A.~Mar\v cenko and L.~A.~Pastur, \tit{Distribution of eigenvalues of some sets of random matrices} Math.~USSR-Sb. {\bf 1}, 457 (1967).
  % in the TDL

  
  
\bibitem{ran_vec} M.~\v Znidari\v c, \tit{Entanglement of random vectors} J.~Phys.~A {\bf 40}, F105 (2007).

\bibitem{foot2} Strictly speaking the equality holds in the leading order in system size because we are replacing the average power with the power of the average, $\ave{\lambda_k}^p \approx \ave{\lambda_k^p}$, however, fluctuations are in almost all cases negligible, $\sigma(I) \ll I$, except at very long times (much longer than $\tc$).

\bibitem{fut} An exception is a random circuit with 2-qubit Haar random gates in the S-OBC configuration where the transition in the purity decay rate on the other hand happens in a window of width $\propto n$~\cite{skin23}.
  
\bibitem{vinayak} Vinayak and M.~\v Znidari\v c, \tit{Subsystem dynamics under random Hamiltonian evolution} J.~Phys.~A {\bf 45}, 125204 (2021).

\bibitem{brandao} F.~G.~S.~L. Brand\~ ao, P.~Cwiklinski, M.~Horodecki, P.~Horodecki, J.~K.~Korbicz, and M.~Mozrzymas, \tit{Convergence to equilibrium under a random Hamiltonian} Phys.~Rev.~E {\bf 86}, 031101 (2012).


  
\bibitem{anto10} A.~{De Pasquale}, P.~Facchi, G.~Parisi, S.~Pascazio, and A.~Scardicchio, \tit{Phase transitions and metastability in the distribution of the bipartite entanglement of a large quantum system} Phys.~Rev.~A {\bf 81}, 052324 (2010).
  % beta(temperature)-ensemble: Boltzmannian beta-weight to the random-state distribution

\bibitem{Nadal11} C.~Nadal, S.~N.~Majumdar, and M.~Vergassola, \tit{Statistical distribution of quantum entanglement for a random bipartite state} J.~Stat.~Phys. {\bf 142}, 403 (2011).
% cela porazdelitev npr. I, kjer imas prehode -- veliki/mali/srednji I imajo drugacno funkc. obliko; bolj splosno kot anto10


\bibitem{jed} P.~Chang, X.~Chen, S.~Gopalakrishnan, and J.~H.~Pixley, \tit{Evolution of entanglement spectra under generic quantum dynamics} Phys.~Rev.~Lett. {\bf 123}, 190602 (2019).


  
  
\end{thebibliography}
\end{document}